\documentclass[12pt]{article}
\usepackage{amsmath}
\usepackage{amssymb}
\usepackage{graphicx}
\usepackage{psfrag}
\usepackage{epsf}
\usepackage{enumerate}
\usepackage{natbib}
\usepackage{url} 
\usepackage{lscape}
\usepackage{booktabs}
\usepackage{setspace}
\usepackage[margin=2.5cm]{geometry}
\newcommand{\E}{\mathbb{E}}
\newcommand{\bell}{\boldsymbol{\ell}}
\newcommand{\bO}{\textbf{\textit{O}}}
\newcommand{\bW}{\textbf{\textit{W}}}
\newcommand{\bbeta}{\boldsymbol{\beta}}
\newcommand{\btheta}{\boldsymbol{\theta}}
\newcommand{\balpha}{\boldsymbol{\alpha}}

\newcommand{\bgamma}{\boldsymbol{\gamma}}

\newcommand{\Var}{\text{Var}}
\newcommand{\logit}{\text{logit}}
\newcommand{\mY}{\mathcal{Y}}
\newcommand{\bL}{\textbf{\textit{L}}}
\newcommand{\ci}{\mathrel{\text{\scalebox{1.07}{$\perp\mkern-10mu\perp$}}}}
\newtheorem{algor}{Algorithm}

\begin{document}

\def\spacingset#1{\renewcommand{\baselinestretch}%
{#1}\small\normalsize} \spacingset{1}

\vspace*{0.4in}

\begin{center}
\begin{doublespace}
{\LARGE  A causal approach to analysis of censored medical costs in the presence of time-varying treatment}

\vspace{0.2in}

{\normalsize Andrew J. Spieker$^{*}$, Arman Oganisian$^{*}$, Emily M. Ko$^{**}$, Jason A. Roy$^{*}$, and Nandita Mitra$^{*}$}

{\normalsize $^{*}$University of Pennsylvania, U.S.A.}

\vspace{-0.1in}

{\normalsize $^{**}$Pennsylvania Hospital, U.S.A.}
\end{doublespace}
\end{center}

\vspace{-0.2in}

\begin{abstract}
There has recently been a growing interest in the development of statistical methods to compare medical costs between treatment groups. When cumulative cost is the outcome of interest, right-censoring poses the challenge of informative missingness due to heterogeneity in the rates of cost accumulation across subjects. Existing approaches seeking to address the challenge of informative cost trajectories typically rely on inverse probability weighting and target a net ``intent-to-treat" effect. However, no approaches capable of handling time-dependent treatment and confounding in this setting have been developed to date. A method to estimate the joint causal effect of a treatment regime on cost would be of value to inform public policy when comparing interventions. In this paper, we develop a nested g-computation approach to cost analysis in order to accommodate time-dependent treatment and repeated outcome measures. We demonstrate that our procedure is reasonably robust to departures from its distributional assumptions and can provide unique insights into fundamental differences in average cost across time-dependent treatment regimes.
\end{abstract}

\section{Introduction}
\label{s:intro}

Rising medical costs are becoming an increasingly important factor in choosing between comparably safe and effective drugs or therapies. In a given disease setting, studies to compare mean cumulative costs across treatment groups are therefore of high value for the purposes of making informed choices. In the ideal case, such a study would include data on lifetime medical costs after commencing treatment, though this is typically not realistic due to time and budget constraints. Prior studies have therefore identified cumulative cost over some fixed time interval after commencing treatment as a suitable outcome measure.

\begin{figure}[h!]
\centering
\includegraphics[width = 5.6in]{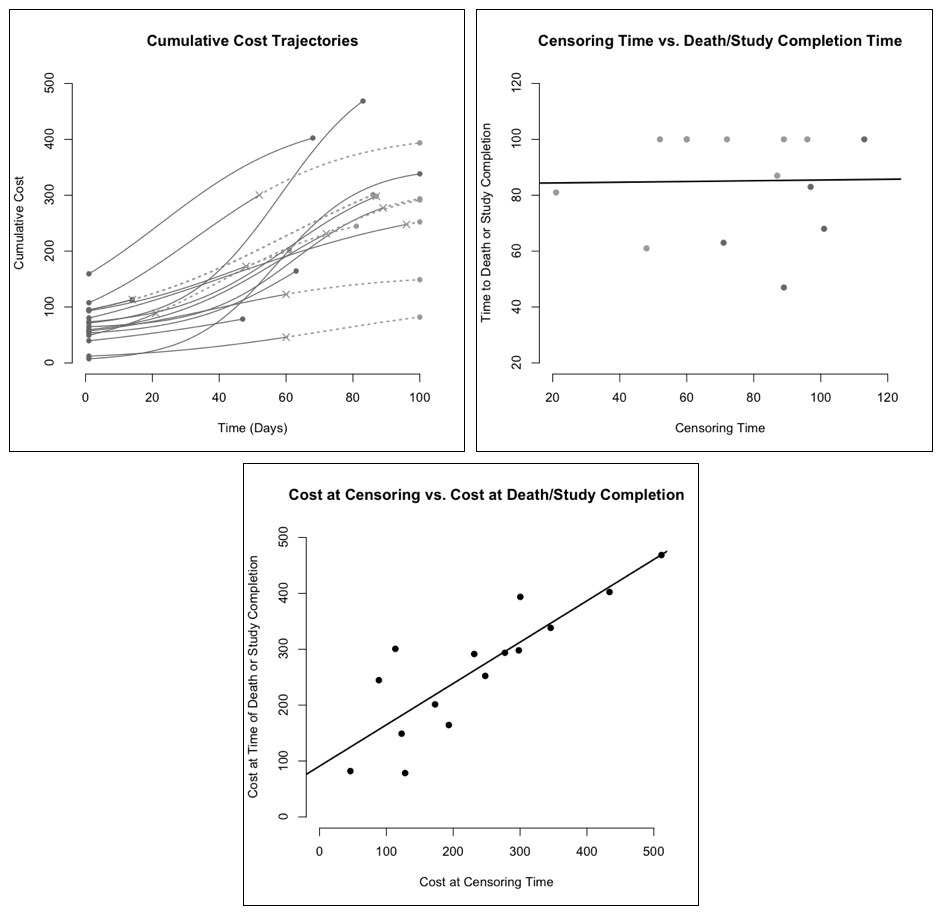}
\caption{
Illustration of the ``informative cost trajectories" phenomenon using a small simulated data set ($N = 15$), with one-hundred-day cumulative cost outcomes and independent censoring and death times; in the upper left panel, observed cost trajectories (solid dark) are shown until either death/study completion ($\bullet$) or censoring ($\times$). For censored individuals, the theoretical remainder of the trajectory to death or study completion is depicted (dashed). Though the censoring and study completion times are themselves independent (upper right panel: light circles indicate complete cases, and dark circles indicate censoring), the cost at censoring is highly correlated with the cost at death (lower panel).
}
\label{F1}
\end{figure}

Even when focusing on costs over some fixed period time, it is highly likely that censoring will occur in some participants; when participants' survival times are censored, so too are their cumulative cost outcomes. Straightforward approaches such as the Kaplan-Meier method and the Cox Proportional Hazards model, despite their utility when seeking to compare survival distributions in the presence of right-censoring, are not appropriate when the outcome of interest is cumulative medical cost. Due to the heterogeneous nature of cost accumulation across subjects, the total cost at the time of censoring is generally not independent of the theoretical total cost over the fixed time interval of interest, even if the censoring time is itself completely independent of the time to death or study completion (illustrated in Figure \ref{F1}). Analyzing medical cost as a time-to-event measure is therefore a fundamentally false strategy. To avoid confusion, we use the term \textit{informative cost trajectories} to refer to this challenge, and restrict use of the phrase ``informative censoring" to refer to survival times. While informative censoring of the survival times may or may not pose its own set of challenges in studies involving right-censored medical costs, the problem of informative cost trajectories necessarily poses a barrier and must always be addressed.

The development of methods to handle incomplete cost data has been of growing interest in recent years. \citet{Lin97} proposed an estimator for mean costs, expressed as a sum of products of the Kaplan-Meier estimator for death/survival and suitable estimators for average cost within defined time intervals. \citet{Bang00} altered this estimator such that desirable asymptotic properties could be achieved under weaker assumptions, whereby mean costs are estimated within discrete intervals using inverse probability-of-censoring weights based on the Kaplan-Meier estimator. This method was then broadened to compare mean costs across predictors of interest, such as treatment (\citealp{Lin00}; \citealp{Lin03}). \citet{Li16} proposed the inclusion of inverse probability-of-treatment weights (IPTW) as one of several ways to address confounding, and further improved upon existing methods by adopting a Super-Learner algorithm to accommodate more flexible cost distributions.

Existing approaches to compare costs are based on intent-to-treat (ITT) analyses, identifying the net causal effect of baseline treatment on cost irrespective of any post-baseline modifications to treatment. However, participants may sometimes switch treatments throughout the course of a study in a manner related to prior observed treatment toxicity, improvements in health, changes to insurance, or even prohibitively high prior medical costs. Time-varying treatment for any of these reasons obscures our ability to directly attribute cost contrasts to different treatments.  Though results from ITT analyses are uniquely equipped to provide insights into the real-world effectiveness of a treatment or therapy on a clinical outcome, an approach to estimate population-average causal effects would likely be of great value for informing policy makers when the outcome is total medical cost.

Another limitation of existing approaches is that they generally presume that the probability of censoring is not informative of death times, conditional on measured covariates at baseline. Existing approaches can accommodate, for example, a study with staggered entry in which the distribution of baseline confounders is related to study enrollment time. They have not, however, permitted censoring at a given time to be related to cost accumulated up until that point, or treatment/confounding history. Particularly in the presence of heavy censoring due to subject withdrawal (non-administrative in nature), strict, untestable assumptions regarding the censoring mechanism can be a serious limitation to the interpretability of results.

The goal of this work is to develop an approach to cost analysis that accommodates time-varying treatment and confounding. G-computation has gained attention as an approach to compare mean outcomes under multiple hypothetical treatment regimes (\citealp{Robins86}). We therefore consider this framework as an alternative to existing approaches in order to assess joint causal effects. As will be made apparent, it will be necessary to modify the standard g-formula to allow reasonable assumptions regarding censoring and death.

The remainder of this paper is organized as follows. In Section 2, we provide a brief summary of the standard g-computation procedure (which would be appropriate for cost analysis in the total absence of censoring and death). In Section 3, we present our \textit{nested} g-computation approach in which the g-formula is iteratively applied to repeated cost outcomes. In Section 4, we empirically compare the nested g-formula to existing approaches and evaluate the sensitivity of our model to departures from its assumptions. We will conclude with a discussion of our findings, including study limitations and potential future research directions.

\section{A brief review of g-computation}

In this section, we briefly describe how the g-formula could be applied to cost data in the absence of censoring and death. More specific details are provided by \cite{Robins86}; \citet{Daniel13} provide an excellent tutorial on how to implement g-computation. 

Let $i = 1, \dots, N$ index independently sampled subjects, and let $j = 1, \dots, J$ index equally spaced intervals over the time range of interest: $0 \equiv \tau_0 < \cdots < \tau_J \equiv \tau$. Let $A_j$ and $\bL_j$ denote treatment status and confounders at the start of interval $j$, respectively, and let $\mY$ denote cumulative cost over the interval $[0, \tau]$. We use overbars to denote variable history, dropping subscripts for entire histories (e.g., $\bar{A}_j = (A_1, \dots, A_j)$, and $\bar{A} = (A_1, \dots, A_J)$).

\begin{figure}[h!]
\centering
\includegraphics[width = 5.8in]{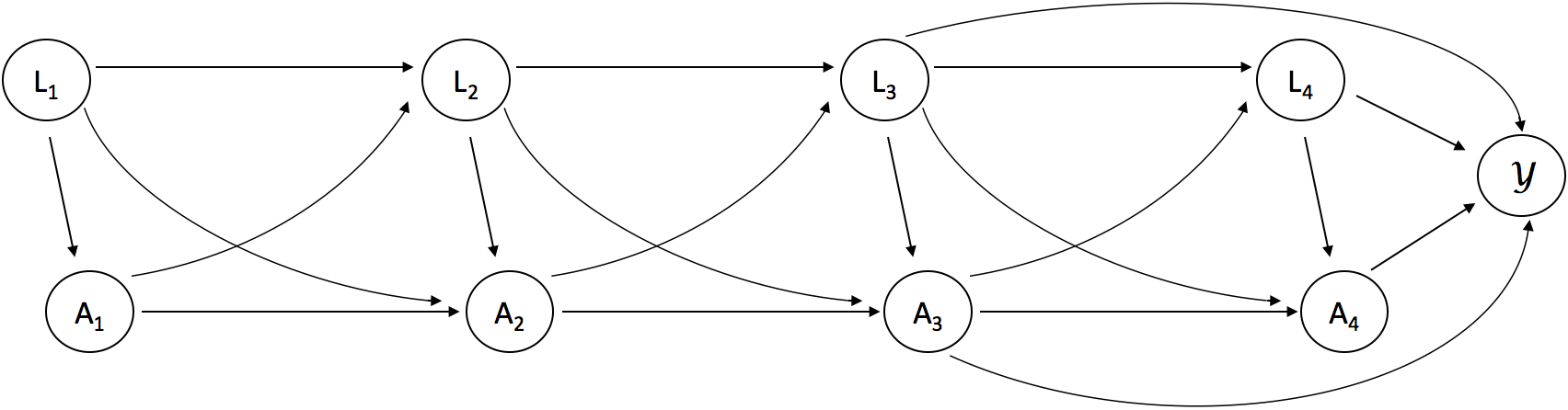}
\caption{Illustration of time-dependent treatment and confounding in the setting of no censoring or death. This figure invokes the Markov assumption that observations from a single interval depend on the prior interval as opposed to all prior intervals.}
\label{F2}
\end{figure}

We adopt the counterfactual notation of \cite{Rubin78}, extended by \citet{Robins86}. Let $\mathcal{A}$ denote the set of all treatment regimes, and $\mY^{\bar{a}}$ the potential cost under treatment $\bar{A} = \bar{a}$ ($\bar{a} \in \mathcal{A}$). Then, $\E[\mY^{\bar{a}}]$ is the expected cost in the hypothetical case in which everyone receives treatment $\bar{A} = \bar{a}$. Letting $\mathcal{L}$ denote all values of $\bar{\bL}$, and $\bar{\bell} \in \mathcal{L}$, the g-formula provides an expression for $\E[\mY^{\bar{a}}]$ by standardizing in the appropriate temporal order:
\begin{eqnarray}
\E[\mY^{\bar{a}}] & = & \int_{\bar{\bell} \in \mathcal{L}} \E[\mY|\bar{\bL} = \bar{\bell}, \bar{A} = \bar{a}]\prod_{j = 1}^{J}f_{\bL_j|\bar{\bL}_{j - 1}, \bar{A}_{j - 1}}(\bell_j, \bar{\bell}_{j - 1}, \bar{a}_{j - 1}) d\bar{\bell},
\end{eqnarray}
where $f(\cdot)$ denotes the density function. This formula relies on the following assumptions:
\begin{enumerate}
\item Stable unit treatment value assumption (SUTVA): $\mY_i^{\bar{a}} \ci \bar{A}_{i'}$, $1 \leq i \neq i' \leq N$ (the potential cost is not influenced by the treatment assignment of others).
\item Consistency: $\mY_{i} = \sum_{\bar{a} \in \bar{\mathcal{A}}}\mY_{i}^{\bar{a}}1(\bar{A}_{i} = \bar{a})$ for $1 \leq j \leq J$ (the observed cost is equal to the potential cost under treatment $\bar{A}_{i} = \bar{a}$).
\item Positivity: $0 < P(A_{ij} = 1|\bar{\bL}_{ij}, \bar{A}_{i(j - 1)}) < 1$, $1 \leq j \leq J$ (each potential treatment regime has nonzero probability of occurrence irrespective of covariate history).
\item Sequentially ignorable treatment assignment: $\mY_{i}^{\bar{a}} \ci A_{ij}|\bar{\bL}_{ij}, \bar{A}_{i(j - 1)}$, $1 \leq j \leq J$ (conditional on prior variables, treatment is independent of the potential cost).
\end{enumerate}
The process by which the integral in Equation (1) is estimated is known as \textit{g-computation}. As it is written, one would typically need to model $\E[\mY|\bar{\bL}, \bar{A}]$ and $f(\bL_{j}|\bar{\bL}_{j - 1}, \bar{A}_{j - 1})$ in order to evaluate it. In settings where $\bL$ contains continuous predictors, it is generally necessary to parametrically model $\E[\mY|\bar{\bL}, \bar{A}]$. In this instance, Equation (1) can be written as:
\begin{eqnarray}
\E[\mY^{\bar{a}}] & = & \int_{\bar{\bell} \in \mathcal{L}} \int_{y \in \text{supp}(\mathcal{Y}|\bar{\bL}, \bar{A})} y f_{\mY|\bar{\bL}, \bar{A}}(y, \bar{\bell}, \bar{a})\prod_{j = 1}^{J}f_{\bL_j|\bar{\bL}_{j - 1}, \bar{A}_{j - 1}}(\bell_j, \bar{\bell}_{j - 1}, \bar{a}_{j - 1}) dy d\bar{\bell},
\end{eqnarray}
where $\text{supp}(\mathcal{Y}|\bar{\bL}, \bar{A}) = \lbrace y : f_\mathcal{Y|\bar{\bL}, \bar{A}}(y, \bar{\bell}, \bar{a}) \neq 0 \rbrace$. Evaluation of this integral typically requires Monte-Carlo integration, whereby data are repeatedly simulated under estimated model parameters and then averaged. To compare mean costs, one can plug in comparator hypothetical treatment regimes into the g-formula and define suitable contrasts (e.g., differences) that characterize joint causal effects. Invoking the Markov assumption (Figure \ref{F2}) can reduce computational burden, and the nonparametric bootstrap can be used to conduct inference.

\section{The nested g-formula for cost estimation}

In settings where complete cost data are available on all subjects, the approach described in Section 2 is valid. When censoring and death events occur, however, the assumptions put forth in the g-computation framework do not naturally generalize in any apparent way. For example, it is not reasonable to assume that censoring is completely independent of the cost outcome, but it is not clear how to formulate a conditional independence assumption on the basis of observed variables.

In this section, describe how repeated cost measures can be used to formulate reasonable assumptions regarding censoring and death. We then present the nested g-formula as a way to estimate marginal mean costs and describe in detail how to implement the associated nested g-computation procedure.

\subsection{Repeated cost measures: Time-updating censoring/death risks}

Let $Y_{j}$ denote cost accumulated in interval $j$, and $\bar{Y}_{j}^{\bar{a}}$ the potential cost in interval $j$ under treatment $\bar{A} = \bar{a}$, noting that $\mY_i^{\bar{a}} = \sum_{j = 1}^{J} Y_{ij}^{\bar{a}}$. Let $C_j$ denote the indicator of censoring at the start of interval $j$, and $D_j$ the indicator of death following interval $j$. We assume that $D_{ij} = 1 \Rightarrow Y_{ij'} = 0$ for $j < j' \leq J$ (that is, no cost is accumulated after death). In this sense, subjects who die prior to their censoring times have complete cost data.

\begin{figure}[h!]
\centering
\includegraphics[width = 5.8in]{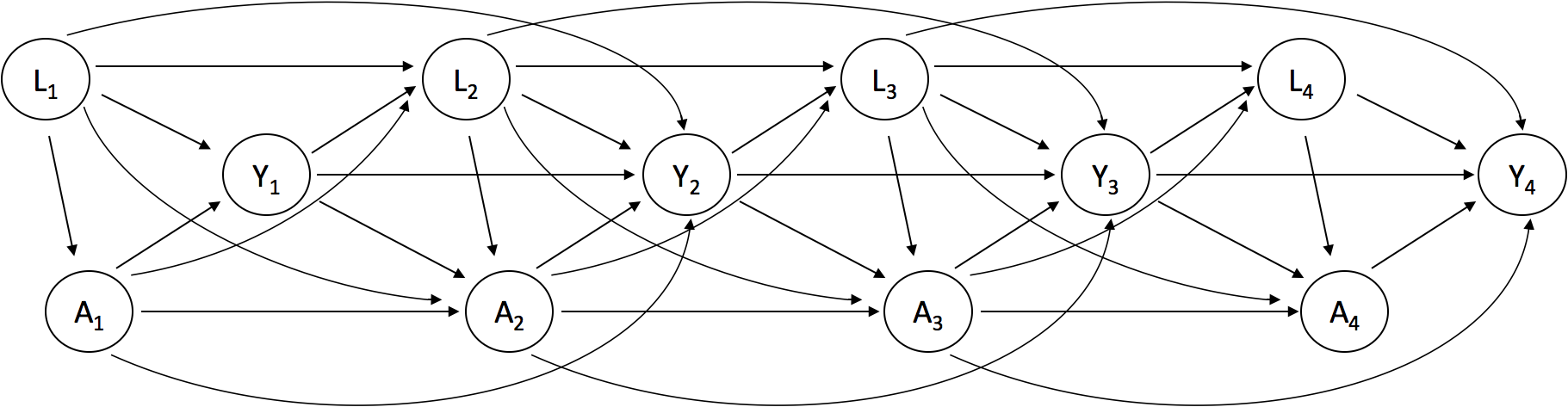}
\caption{
Illustration of time-dependent treatment and confounding in the setting where repeated cost measures are available. This figure invokes the Markov assumption that observations from a single interval depend on the one prior and not on all prior intervals.
}
\label{F3}
\end{figure}

The concept of a complete treatment regime is only meaningful when taken in the absence of censoring; we take this as implicit in our notation, though the mean total coast under treatment history $\bar{A} = \bar{a}$ could be more precisely denoted as $\E[\mY^{\bar{a}, C_{J} = 0}]$. In the setting of repeated cost measures, we update the assumptions listed in Section 2:

\begin{enumerate}
\item Stable unit treatment value assumption (SUTVA): $Y_{ij}^{\bar{a}} \ci \bar{A}_{i'}$, $1 \leq i \neq i' \leq N$ (the potential cost history is not influenced by the treatment assignment of others).
\item Sequential consistency: $\bar{Y}_{i} = \sum_{\bar{a} \in \bar{\mathcal{A}}}\bar{Y}_{i}^{\bar{a}}1(\bar{A}_{i} = \bar{a})$ for $1 \leq j \leq J$ (the observed cost history is equal to the potential cost history under treatment $\bar{A}_{i} = \bar{a}$).
\item Positivity: $0 < P(A_{ij} = 1|\bar{\bL}_{ij}, \bar{A}_{i(j - 1)}, \bar{Y}_{i(j - 1)}) < 1$, $1 \leq j \leq J$ (each potential treatment regime has nonzero probability of occurrence irrespective of covariate history).
\item Sequentially ignorable treatment assignment: $Y_{ij}^{\bar{a}} \ci A_{ij}|\bar{\bL}_{ij}, \bar{A}_{i(j - 1)}, \bar{Y}_{i(j - 1)}$, $1 \leq j \leq J$ (conditional on previously observed variables, treatment status is independent of the potential costs in each interval).
\item Sequentially ignorable censoring: $Y_{ij}^{\bar{a}} \ci C_{ij}|\bar{\bL}_{ij}, \bar{A}_{ij}, \bar{Y}_{i(j - 1)}$, $1 \leq j \leq J$.
\item Sequentially ignorable death: $Y_{ij}^{\bar{a}} \ci D_{i(j - 1)}|\bar{\bL}_{ij}, \bar{A}_{ij}, \bar{Y}_{i(j - 1)}, C_{i(j - 1)}$, $1 \leq j \leq J$.
\item Sequentially non-informative censoring: $C_{ij} \ci D_{i(j - 1)}|\bar{\bL}_{ij}, \bar{A}_{ij}, \bar{Y}_{ij}$, $1 \leq j \leq J$.
\end{enumerate}
We refer to the second assumption as \textit{sequential} consistency to highlight the use of repeated outcomes; the consistency assumption expressed in terms of $\mY^{\bar{a}}$ would not be sufficient. Assumptions 5 and 6 state that conditional on observed variable history, censoring and death are independent of the potential cost. Assumption 7 essentially states that two individuals of the same variable history, differing only in their censoring status at the beginning of an interval, are equally likely to die in that interval. We assume the following temporal ordering of variables within an interval: $(C_j, \bL_j, A_j, Y_j, D_j)$. Figure \ref{F3} depicts time-dependent treatment and confounding in the setting where repeated cost measures are available.

\subsection{The nested g-formula}

We seek to construct a formula for the mean cumulative cost over under some hypothetical treatment regime, $\bar{A} = \bar{a}$. This entails applying the g-formula to variables through each interval to estimate $\E[Y_j^{\bar{a}}]$, from which an aggregated mean can be formed by invoking linearity of expectation: $\E[\mY^{\bar{a}}] = \sum_{j = 1}^{J}\E[Y_j^{\bar{a}}]$. Under the assumptions of Section 3.1, we have that
\begin{eqnarray*}
\E[\mY^{\bar{a}}] &=& \sum_{j = 1}^{J} \iiint \left\lbrace y_j  \prod_{k = 1}^{j} f_{Y_k|\bar{Y}_{k - 1}, \bar{C}_k, \bar{\bL}_k, \bar{A}_k, \bar{D}_{k - 1}} (y_k, \bar{y}_{k - 1}, 0, \bar{\bell}_k, \bar{a}_k, \bar{d}_{k - 1}) \right. \\
~ & ~ & \hspace{0.8in} \times \hspace{0.03in} \prod_{k = 1}^{j} f_{D_k|\bar{D}_{k - 1}, \bar{Y}_{k}, \bar{C}_{k}, \bar{\bL}_{k}, \bar{A}_{k}} (d_k, \bar{d}_{k - 1}, \bar{y}_{k}, 0, \bar{\bell}_{k}, \bar{a}_{k}) \\
~ & ~ & \hspace{0.8in} \times \hspace{0.01in} \left. \prod_{k = 1}^{j} f_{L_k|\bar{\bL}_{k - 1}, \bar{A}_{k - 1}, \bar{D}_{k - 1}, \bar{Y}_{k - 1}, \bar{C}_{k}} (\bell_k, \bar{\bell}_{k - 1}, \bar{a}_{k - 1}, \bar{d}_{k - 1}, \bar{y}_{k - 1}, 0) \right\rbrace d\bar{d}_{j} d\bar{y}_{j} d\bar{\bell}_j.
\end{eqnarray*}

Appendix A presents the derivation of the nested g-formula. The suppressed limits of integration are analogous to those in Section 2. By convention, $D_k = 1 \Rightarrow L_{k'} = A_{k'} = Y_{k'} = 0$, and $D_{k'} = 1$ for $k' > k$. Writing the densities in the integrand as conditional on not have been censored is implicit in the sense that these are requirements to be observed. Without Assumptions 5 through 7, densities could not be expressed on the basis of observed data. The nested g-formula expresses the total mean cost as the sum of mean interval costs over each interval. By linearity of expectation, this is equivalent to sequentially iterating the standard g-formula, treating each repeated outcome as a confounder for future observations. 

This novel approach has clear advantages. Challenges surrounding informative censoring of survival times are addressed by allowing time-updating censoring and death risks according to observed variables. These assumptions allow for censoring to be handled similarly to treatment in that censoring values are set to zero in the g-formula. Sequentially conditioning each cost outcome on cost history helps reconstruct the heterogeneity in cost trajectories, while allowing subjects to contribute to estimation of marginal means for as long as they are not censored. Repeated outcome measures have previously been used in the setting of discrete time-to-event problems in order to evaluate cumulative risk (\citealp{Taubman09}); however, our problem fundamentally differs in that each of the continuous outcomes sequentially impacts future variables and need not be monotonically increasing.

\subsection{Nested g-computation for comparing marginal means}

We now describe how the nested g-computation procedure can be applied to estimate (and in turn, compare) mean costs. We specifically consider the case in which $\bL$ contains continuous variables. Then, parametric models would be appropriate for: the cost distribution ($Y_j|Y_{kj- 1}, \bar{C}_{j} = 0, \bar{\bL}_j, \bar{A}_j, \bar{D}_{j - 1} = 0$), the death risk ($D_j|\bar{D}_{j - 1} = 0, \bar{Y}_{j}, \bar{C}_{j} = 0, \bar{\bL}_{j}, \bar{A}_{j}$), and the confounder distribution ($\bL_j|\bar{\bL}_{j - 1}, \bar{A}_{j - 1}, \bar{D}_{j - 1} = 0, \bar{Y}_{j - 1}, \bar{C}_{j} = 0$).  Under the Markov assumption, one could first estimate baseline parameters. Let $\btheta_1 = (\balpha_1, \bbeta_1, \bgamma_1)$ index the confounder, cost, and death models in the first interval, respectively. The analogous post-baseline parameters, $\btheta = (\balpha, \bbeta, \bgamma)$, can be estimated using all available post-baseline data. Maximum partial likelihood is one way to estimate $(\btheta_1, \btheta)$ from some user-specified score equations. For example, the cost model score equations would be given by:
\begin{eqnarray*}
\sum_{i = 1}^{N} \frac{\partial}{\partial \bbeta_1} \log f(Y_{i1}|C_{i1} = 0, \bL_{i1}, A_{i1}; \bbeta_1) & = & \textbf{0}; \\
\sum_{i = 1}^{N} \sum_{j = 2}^{J_i}\frac{\partial}{\partial \bbeta} \log f(Y_{ij}|\bar{C}_{ij} = 0, \bL_{i(j - 1)}, A_{i(j - 1)}, Y_{i(j - 1)}, D_{i(j - 1)} = 0, \bL_{ij}, A_{ij}; \bbeta) & = & \textbf{0},
\end{eqnarray*}
where $J_i$ denotes the number of observations for subject $i$. Analogous equations exist for the confounder and death models. Model specification can be made somewhat complex, although model selection is not the focus of this paper. Death could be modeled using, for example, logistic regression, and cost could be modeled using a generalized linear model (GLM). The normal GLM with an identity link would amount to performing ordinary least squares, although the Inverse Gaussian (Inv-$\mathcal{N}$) and Gamma distributions may be appropriate when cost is right-skewed, and are appealing due to their support on the positive real line.

As described in the setting of the standard g-formula in Section 2, the high-dimensional integral in the nested g-formula generally does not possess a closed-form expression. Given estimates $(\hat{\btheta}_1, \hat{\btheta})$, a Monte-Carlo procedure to obtain an estimate (say, $\hat{\mu}_{\mY^{\bar{a}}}$) of $\E[\mY^{\bar{a}}]$ for some treatment regime of interest is given as follows.
\begin{algor}
Monte-Carlo Integration for Nested G-Formula
\begin{enumerate}
\item Set $\bar{A} = \bar{a}$. For $r = 1, \dots, R$ (a sufficiently large number of Monte-Carlo iterations), generate random draws as follows on the basis of estimated model parameters:
\begin{itemize}
\item $\bL_{1}^r|\hat{\balpha}_1$
\item $Y_{1}^r|(L_{1}^r, A_1 = a_1; \hat{\bbeta}_1)$
\item $D_{1}^r|(L_{1}^r, A_1 = a_1, Y_1^r; \hat{\bgamma}_1)$
\item For $j = 2, \dots, J$
\begin{itemize}
\item $\bL_{j}^r|(\bL_{j - 1}^r, A_{j - 1} = a_{j - 1}, Y_{j - 1}^r, D_{j - 1}^r = 0; \hat{\balpha})$
\item $Y_{j}^r|(\bL_{j - 1}^r, A_{j - 1} = a_{j - 1}, Y_{j - 1}^r, D_{j - 1}^r = 0, \bL_{j}^r, A_{j} = a_{j}; \hat{\bbeta})$
\item $D_{j}^r|(\bL_{j - 1}^r, A_{j - 1} = a_{j - 1}, Y_{j - 1}^r, D_{j - 1}^r = 0, \bL_{j}^r, A_{j} = a_j, Y_{j}^r; \hat{\bgamma})$
\end{itemize}
\item Set $\mY_{r} = \sum_{j = 1}^{J} Y_{j}^r$
\end{itemize}
\item $\E[\mY^{\bar{a}}]$ can be estimated with $\hat{\mu}_{\mY^{\bar{a}}} = R^{-1}\sum_{r = 1}^{R} \mY_{r}$.
\end{enumerate}
\end{algor}
Importantly, the $J$ expectations that comprise the nested g-formula need not be represented by $J$ separate Monte-Carlo procedures; complete data can be generated for from which mean interval costs can be summed. Typically, a study to compare costs targets a contrast (e.g., a difference) in means. Define $\Delta = \E[\mY^{\bar{a}}] - \E[\mY^{\bar{a}'}]$ for some comparator treatment regimes. In turn, $\widehat{\Delta} = \hat{\mu}_{\mY^{\bar{a}}} - \hat{\mu}_{\mY^{\bar{a}'}}$ can be estimated by applying Algorithm 1 to $\bar{a}$ and $\bar{a}'$.

\subsection{The nonparametric bootstrap for inference regarding $\Delta$}

There is generally no general closed-form analytic expression to estimate the repeat-sample variance of $\widehat{\Delta}$. The nonparametric bootstrap, commonly implemented in g-computation procedures, can therefore be used to estimate $\Var[\widehat{\Delta}]$. This procedure is described as follows:

\begin{algor}
The Nonparametric Bootstrap for Estimation of $\Var[\widehat{\Delta}]$
\begin{enumerate}
\item For $b = 1, \dots, B$ (a sufficiently large number of bootstrap replicates)
\begin{itemize}
\item Form dataset ``$b$" by resampling $N$ subjects with replacement from observed data.
\item Obtain estimates $(\hat{\btheta}_1^b, \hat{\btheta}^b)$ on the basis of data set ``$b$"
\item Apply Algorithm 1 to $\bar{A} = \bar{a}$ and then to $\bar{A} = \bar{a}'$ on the basis of $(\hat{\btheta}_1^b, \hat{\btheta}^b)$; take the difference of the results to obtain the bootstrap realization $\widehat{\Delta}_b$
\end{itemize}
\item Compute $\overline{\Delta} = B^{-1}\sum_{b = 1}^{B} \widehat{\Delta}_b$
\item An estimate can be obtained as follows: $\widehat{\Var}[\widehat{\Delta}] = (B - 1)^{-1}\sum_{b = 1}^{B} (\widehat{\Delta}_b - \overline{\Delta})^2$
\end{enumerate}
\end{algor}

Wald-based confidence intervals can be formed on the basis of this variance estimator and inference can be conducted by comparing the quantity $\widehat{\Var}[\widehat{\Delta}]^{-1/2}(\widehat{\Delta} - \Delta_0)$ to the appropriate quantiles of standard normal distribution for some null value $\Delta_0$ (e.g., zero under the null hypothesis of no difference in total mean cost). Other more sophisticated modifications to the standard nonparametric bootstrap approach are described by \citet{Davison97}.

\section{Simulation studies}
\label{s:sims}

In this section, we conduct a number of simulation studies in order to empirically assess the performance of our nested g-formula for cost estimation. We are specifically interested in (i) evaluating our approach in terms of bias and variability in the presence of various forms of censoring, (ii) understanding the sensitivity of our approach to departures from distributional assumptions on repeated cost, and (iii) comparing the nested g-formula to the previously developed inverse-weighting approaches.

\subsection{General simulation setup}

Suppose $N =$ 1,000 subjects, each with a maximum of $J = 6$ observations. We utilize $R =$ 100,000 Monte-Carlo iterations for the g-formula and $B = 100$ bootstrap samples to compute standard errors. We assume a binary treatment. In all cases, we invoke Markov assumption of Figure \ref{F3}. Let $\bO_j = (L_{j}, A_{j}, Y_{j})$, for ease of notation. To focus attention on the primary study goals, we hold fixed a number of simulation characteristics.

All participants are presumed not to be censored at the start of the study ($C_1 = 0$). We generate a single normally distributed time-varying confounder (e.g., some appropriately transformed propensity score, as per \citet{Lu05}), This is generated as $L_1 \sim \mathcal{N}(\alpha_{0,1}, \sigma_{L_1}^2)$ at baseline, and $L_j|(D_{j - 1} = C_{j} = 0, \bO_{j - 1}) \sim \mathcal{N}(\alpha_0 + \alpha_1L_{j - 1} + \alpha_2A_{j - 1} + \alpha_3Y_{j - 1}, \sigma_{L}^2)$ at follow-up. The odds of treatment ``1" in the first interval is given by $\logit\lbrace P(A_1 = 1|L_1)\rbrace = \eta_{0,1} + \eta_{1,1} L_1$, and at follow-up intervals by
\begin{eqnarray*}
\logit \lbrace P(A_j = 1|D_{j - 1} = C_{j} = 0, L_{j - 1}, \bO_{j - 1}, L_{j}) \rbrace &=& \eta_0 + \eta_1 L_{j - 1} + \eta_2 L_j + \eta_3 A_{j - 1} + \eta_4 Y_{j - 1}.
\end{eqnarray*}
We assume comparably safe interventions, so that death is independent of treatment, conditional on prior cost and confounders. The odds of death following the first interval is therefore given by $\logit \lbrace P(D_1 = 1|\bO_1) \rbrace = \gamma_{0,1} + \gamma_{1, 1} L_{1} + \gamma_{2, 1} Y_1$, and the odds of death following each subsequent interval by 
\begin{eqnarray*}
\logit \lbrace P(D_j = 1|D_{j - 1} = C_{j} = 0, \bO_{j - 1}, \bO_j) \rbrace &=& \gamma_0 + \gamma_1 L_{j - 1} + \gamma_2 L_j + \gamma_3 Y_j.
\end{eqnarray*}
Finally, we assume that the mean cost takes the form $\mu_{1} = \beta_{0, 1} + \beta_{1, 1}L_{1} + \beta_{1, 2}A_{1}$ in the first interval, and $\mu_{j} = \beta_{0} + \beta_{1}L_{j - 1} + \beta_{2}L_{j} + \beta_{3} A_{j - 1}+ \beta_{4} A_{j} + \beta_{5} Y_{j - 1}$ at follow-up intervals, conditional on $D_{j - 1} = 0$ (otherwise, $Y_j = 0$). We will consider different parametric forms for cost, which we imagine to be in thousands of dollars. In all settings, the target of inference is given by the joint causal effect, $\Delta = \E[\mY^{\bar{a}}] - \E[\mY^{\bar{a}'}]$, where $\bar{a} = (1, \dots, 1)$ and $\bar{a}' = (0, \dots, 0)$. All simulations were conducted in R version 3.3.1 (2013). Appendix B provides a table summarizing the selection of parameters that remain constant across simulations.

\subsection{Results under different censoring mechanisms}

\begin{figure}[h!]
\centering
\includegraphics[width = 5.4in]{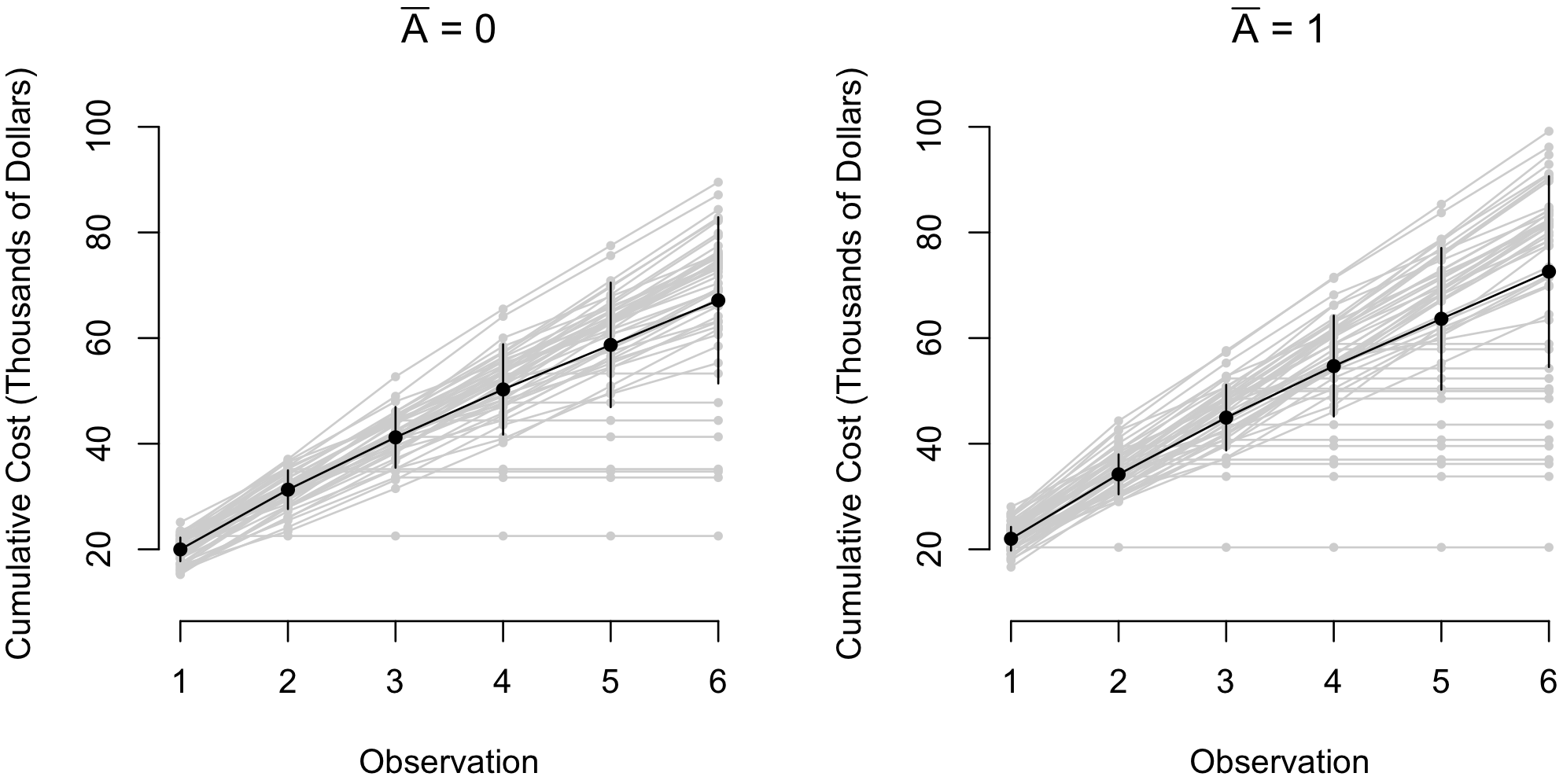}
\caption{Illustration of simulation setup in the absence of censoring. The gray lines represent subject specific trajectories, leveling out once death has occurred. The black lines represent the mean ($\pm$ one standard deviation) cumulative cost at each observation. The heterogeneity in trajectories is apparent in this figure, as seen by the increasing variability over time.}
\label{F4}
\end{figure}

The goal of this simulation is to confirm that the nested g-formula provides valid estimates of $\Delta$ under various forms of censoring. We consider three scenarios, from the least to most informative. In the first scenario, we assume that censoring occurs completely at random: $P(C_{j} = 1|C_{j - 1} = D_{j - 1} = 0, \bO_j) = 0.05$. In the second scenario, we assume staggered enrollment such that the characteristics of the baseline confounder vary over time:
\begin{eqnarray*}
\logit \lbrace P(C_{j} = 1|C_{j - 1} = D_{j - 1} = 0, \bO_1)\rbrace = -3.3 + 0.25 L_1 + 0.5 A_1.
\end{eqnarray*}
In the third scenario, we assume that the probability of censoring is time-varying, dependent upon variables in concurrent intervals:
\begin{eqnarray*}
\logit \lbrace P(C_j = 1|C_{j - 1} = D_{j - 1} = 0, \bO_j)\rbrace = -3.5 + 0.25 L_j + 0.5 A_j + 0.01 Y_j.
\end{eqnarray*}
In each of these settings, the probability of censoring within each interval is approximately 0.05. We generate cost outcomes from a normal distribution with mean $\mu_{j}$ ($1 \leq j \leq J$) as given in Section 4.1, with error variance $\sigma_Y^2 = 2$ at each time. Figure \ref{F4} illustrates the nature of the cost trajectories under this setting in the absence of censoring.  Under this simulation setup, the true value of $\mu_{Y^0} \equiv \E[Y^{\bar{a} = 0}]$ was determined to be 66.6 and $\mu_{Y^1} \equiv \E[Y^{\bar{a} = 1}]$ was determined to be 72.0 (thus, $\Delta = 5.40$). Results are depicted in Table \ref{T2}.

\begin{table}[h!]
\centering
\caption{Results from Simulation Study 1 in which a variety of censoring mechanisms are considered. In this setting, $\Delta = 5.40$ represents the true value. Depicted are the average bias, absolute \% bias, Monte-Carlo standard error, average bootstrap standard error estimate, and estimated coverage probability based on Wald-based confidence intervals.}
\label{T2}
\begin{tabular}{rccccc}
\hline
Description &  Bias($\widehat{\Delta}$) & \% Bias($\widehat{\Delta}$) & MCSE$(\widehat{\Delta})$ & $\widehat{\text{SE}}(\widehat{\Delta})$ & CP \\ \hline
Random censoring & 0.012 & 0.227\% & 0.529 & 0.543 & 0.955 \\
Staggered study entry &  0.0082 & 0.152\% & 0.532 & 0.542 & 0.955 \\
Nonrandom dropout &  0.0052 & 0.096\% & 0.531 & 0.543 & 0.949 \\ \hline
\end{tabular}
\end{table}

The nested g-formula performs well regardless of the censoring mechanism and dependence on baseline/concurrent variables. Bias is low, the nonparametric bootstrap standard errors adequately represent the simulation-based standard errors, and two-sided Wald-based 95\% confidence intervals appear to have proper coverage of the true parameter value.

\subsection{Results under various cost distributions}

The purpose of this study is to examine the behavior of the nested g-formula under various parametric models for cost. We consider three different mechanisms to generate cost data, the first mirroring the prior simulation--namely, $Y_j \sim \mathcal{N}(\mu_{j}, \sigma_Y^2 = 2)$. We also consider the two cases in which cost data are right-skewed: namely, $Y_j \sim \text{Gamma}(\alpha = 8, \beta = 8/\mu_{j})$ with mean $\mu_j$ and variance $\mu_{j}^2/\alpha$, and $Y_j \sim \text{Inv-}\mathcal{N}(\mu = \mu_{j}, \lambda = 40)$, with mean $\mu_{j}$ and variance $\mu_{j}^3/\lambda$. In each of these settings, the shape parameters ($\alpha$ and $\lambda$, respectively) can be estimated as the inverse of the quasi-likelihood based dispersion factor.

In each of these scenarios, we fit the cost model using each of the three associated GLMs with the identity link (exactly one of which is correctly specified in each case). We generate data from the nonrandom dropout mechanism, the most informative among those considered in the prior study. Results are depicted in Table \ref{T3}.

\begin{table}[h!]
\centering
\caption{Results from Simulation Study 2 in which a variety of parametric data generation mechanisms and models are considered. The true value for the average cumulative costs (and the difference) differs slightly depending on how the data are generated. Depicted are the average bias, absolute \% bias, simulated standard error, average bootstrap standard error estimate, and estimated coverage probability based on Wald-based confidence intervals.}
\label{T3}
\begin{tabular}{rccccccccc}
\hline
\multicolumn{10}{c}{\underline{Data Generation Mechanism: Normal}}\\
Model & $\mu_{\mY^0}$ & $\mu_{\mY^1}$ & $\hat{\mu}_{\mY^0}$ & $\hat{\mu}_{\mY^1}$ & Bias($\widehat{\Delta}$) & \% Bias($\widehat{\Delta}$) & MCSE$(\widehat{\Delta})$ & $\widehat{\text{SE}}(\widehat{\Delta})$ & CP \\ \hline
Normal & 66.6 & 72.0 & 66.6 & 72.0 & 0.0052 & 0.096\% & 0.531 & 0.543 & 0.949 \\ 
Gamma & 66.6 & 72.0 & 66.7 & 72.1 & -0.0022 & 0.040\% & 0.586 & 0.556 & 0.940  \\
Inv-$\mathcal{N}$ &  66.6 & 72.0 & 66.6 & 72.0 & 0.0148 & 0.275\% & 0.628 & 0.614 & 0.932 \\ \hline
\multicolumn{10}{c}{\underline{Data Generation Mechanism: Gamma}}\\
Model & $\mu_{\mY^0}$ & $\mu_{\mY^1}$ & $\hat{\mu}_{\mY^0}$ & $\hat{\mu}_{\mY^1}$ & Bias($\widehat{\Delta}$) & \% Bias($\widehat{\Delta}$) & MCSE$(\widehat{\Delta})$ & $\widehat{\text{SE}}(\widehat{\Delta})$ & CP \\ \hline
Normal & 66.5 & 71.9 & 66.5 & 71.9 & 0.0070 & 0.13\% & 0.948 & 1.02 & 0.966 \\
Gamma & 66.5 & 71.9 & 66.5 & 71.9 & 0.0054 & 0.100\% & 1.02 & 1.00 & 0.950\\
Inv-$\mathcal{N}$ & 66.5 & 71.9 & 66.5 & 71.9 & -0.0156 & 0.290\% & 1.03 & 1.00 & 0.938\\ \hline
\multicolumn{10}{c}{\underline{Data Generation Mechanism: Inverse Gaussian}}\\
Model & $\mu_{\mY^0}$ & $\mu_{\mY^1}$ & $\hat{\mu}_{\mY^0}$ & $\hat{\mu}_{\mY^1}$ & Bias($\widehat{\Delta}$) & \% Bias($\widehat{\Delta}$) & MCSE$(\widehat{\Delta})$ & $\widehat{\text{SE}}(\widehat{\Delta})$ & CP \\ \hline
Normal & 66.1 & 71.2 & 66.2 & 71.5 &  0.174  & 3.42\% & 1.67  & 1.69  &  0.959 \\
Gamma & 66.1 & 71.2 & 66.2 & 71.3 & 0.062 & 1.21\% & 1.64 & 1.60 & 0.940 \\
Inv-$\mathcal{N}$ & 66.1 & 71.2 & 66.1 & 71.3 & 0.026 & 0.502\% & 1.64 & 1.57 & 0.945 \\ \hline
\end{tabular}
\end{table}

The nested g-formula performs well under correct cost model specification (with low bias, valid bootstrap standard errors, and proper coverage). When cost is generated from either the Normal or Gamma distribution, the Inverse Gaussian model results in the highest bias for estimating $\Delta$. When cost is generated from the Inverse Gaussian distribution, the Normal model produces the highest bias, and the Gamma model produces slightly less. Given the heavier skewness of the Inverse Gaussian distribution relative to the others, these findings are unsurprising. Bootstrap standard errors appear valid regardless of model misspecification.

It could so happen in practice that each mean of interest is estimated with comparable bias in the same direction, resulting in low bias for the difference. This explains the seemingly fair performance of the nested g-formula in the third scenario of Table \ref{T3} despite misspecification. We therefore consider a situation in which the cost distribution varies between treatment arms (a Normal distribution in the control arm an Inverse Gaussian distribution in the treatment arm). In this case, $\mu_{\mY^0} = 66.1$ and $\mu_{\mY^1} = 72.0$, such that $\Delta = 5.92$. If the Normal GLM is used for the cost model in both arms, we find that that $\hat{\mu}_{\mY^0} = 66.5$ and $\hat{\mu}_{\mY^1} = 71.8$, so that Bias($\widehat{\Delta}$) = $-0.550$ and $\%$ Bias($\widehat{\Delta}$) =  $9.29\%$, appreciably higher than the bias seen in Table \ref{T3}. Particular attention should therefore be paid to the parametric assumptions in each arm, as our target of inference relies on estimation of two quantities. 

\subsection{Comparison to inverse weighting approaches}

We compare the nested g-formula to the confounder-adjusted partitioned estimator (\citealp{Lin00}) and the IPTW approach (\citealp{Li16}). Estimation procedures based on these approaches are outlined in Appendix C. Each of these approaches targets the net effect of baseline treatment on cumulative cost, whereas the nested g-formula targets the joint causal effect treatment regime $\bar{A} = 1$ relative to $\bar{A} = 0$. To highlight this distinction, we denote the ITT effect by $\Delta_{\text{ITT}}$. We compare the average estimates and simulated standard errors for each approach under the three parametric distributions discussed in Section 4.3. We consider the settings of random censoring and nonrandom dropout (the former satisfies the assumptions of existing methods). Results are depicted in Table \ref{T4}.
\begin{table}[h!]
\centering
\caption{Results from Simulation 3 in which our nested g-formula is compared to the approaches of Lin and Li. Depicted are the average estimate and the Monte-Carlo standard errors.}
\label{T4}
\begin{tabular}{rcccccccc}
\hline
~ & \multicolumn{2}{c}{Nested g-formula} & & \multicolumn{2}{c}{Lin (Adjusted)} & & \multicolumn{2}{c}{Li (IPTW)}\\ \cmidrule{2-3} \cmidrule{5-6} \cmidrule{8-9}
Random censoring & $\widehat{\Delta}$ & MCSE($\widehat{\Delta}$) & & $\widehat{\Delta}_{\text{ITT}}$ & MCSE($\widehat{\Delta}_{\text{ITT}}$) &  &$\widehat{\Delta}_{\text{ITT}}$ & MCSE($\widehat{\Delta}_{\text{ITT}}$) \\ \hline
Normal ($\Delta = 5.40$) & 5.41 & 0.53 &  & 2.60 & 1.15 & & 2.57 & 1.23   \\
Gamma ($\Delta = 5.37$) & 5.35 & 1.03 &  & 2.52 & 1.42 & & 2.50 & 1.50  \\
Inv-$\mathcal{N}$ ($\Delta = 5.11$) & 5.14 & 1.65 &  & 2.26 & 1.80 &  &2.24 & 1.88  \\ \hline
~ & \multicolumn{2}{c}{Nested g-formula} & & \multicolumn{2}{c}{Lin (Adjusted)} & & \multicolumn{2}{c}{Li (IPTW)}\\ \cmidrule{2-3} \cmidrule{5-6} \cmidrule{8-9}
Nonrandom dropout & $\widehat{\Delta}$ & MCSE($\widehat{\Delta}$) & & $\widehat{\Delta}_{\text{ITT}}$ & MCSE($\widehat{\Delta}_{\text{ITT}}$) &  &$\widehat{\Delta}_{\text{ITT}}$ & MCSE($\widehat{\Delta}_{\text{ITT}}$) \\ \hline
Normal ($\Delta = 5.40$) & 5.40 & 0.53 &     & 2.47 & 1.14 &     & 2.45 & 1.22   \\
Gamma ($\Delta = 5.37$) & 5.38 & 1.02 &  & 2.38 & 1.40 & & 2.37 & 1.48  \\
Inv-$\mathcal{N}$ ($\Delta = 5.11$) & 5.13 & 1.64 &  & 2.12 & 1.78 &  & 2.11 & 1.86  \\ \hline
\end{tabular}
\end{table}

In the setting of random censoring, the ITT estimators do not provide consistent estimates of $\Delta$, but rather the ITT parameter $\Delta_{\text{ITT}}$, which in this setting is much smaller than $\Delta$. The ITT approaches do not accommodate a time-varying censoring mechanism related to cost, as seen in that the estimates are attenuated in the setting of nonrandom dropout. Results from the ITT approaches may therefore provide misleading results about the joint effect of treatment on cumulative cost if used in the context of time-dependent treatment. Moreover, the nested g-formula provides markedly smaller standard errors than the ITT approaches, likely attributable to their complete use of available data. This is not to discredit the merits of the inverse-weighting approaches in estimating the ITT effect, but rather to illustrate that a different target of inference translates to a potentially meaningful difference in conclusions.

\subsection{Wald tests of the joint causal effect}

A test of $H_0 : \Delta = 0$ vs. $H_1 : \Delta \neq 0$ in the context of the nested g-formula is a test of the joint effect of a treatment regime $\bar{a}$ on cumulative cost (relative to some comparator treatment regime $\bar{a}'$). The joint effect is a simultaneous collection of treatment effects, each unmediated by treatment assignment history. There are many ways in which data can be generated under $H_0$, although a simple way would be to set $\beta_{1,2} = \beta_3 = \alpha_2 = 0$ in the context of our simulation setup. In particular, the condition $\beta_{1,2} = \beta_3 = 0$ is insufficient to generate data from the null, as effects of treatment on cost can be mediated by future values of the time-varying confounder (leading to a phenomenon known as the g-null paradox, presented by \citet{Robins86}, and discussed further by \citet{Robins97}). It is not straightforward to express a general null hypothesis in terms of the partial likelihood parameters.

To empirically examine the behavior of the Wald test based on the nonparametric bootstrap standard errors, we generate data under the null hypothesis for each of the three parametric models for cost considered. We estimated the level of the Wald-based bootstrap hypothesis test (at the nominal 5\% level) described in Section 3.4. We obtained an estimated a level of 0.047 under the normal model; under the Gamma and Inverse-Gaussian models, we obtained estimated levels of 0.048 and 0.051, respectively.

\section{Discussion}
\label{s:discuss}

The challenge of informative cost trajectories poses a major barrier to analyzing cost data in the presence of censoring, a fact that has been documented and described in multiple papers (\citealp{Lin97}; \citealp{Bang00}; \citealp{Lin00}; \citealp{Lin03}; \citealp{Li16}). Since time accumulates at the same rate for all individuals, one is not confronted with this fundamental challenge when the outcome of interest is a true time-to-event measure. The monotone increasing nature of cost, together with the heterogeneity in cost accrual rates renders standard approaches for right-censored data invalid. Prior work has primarily focused on inverse probability weighting in order to address informative cost trajectories in a way that is consistent with either time-stable treatment or ITT analyses. In this paper, we have developed an alternative framework to estimate the joint causal effect of a time-varying treatment on mean cumulative medical costs over some interval of interest.

Our approach shares a similarity with existing approaches in that a time partition is desirable in some sense. The nested g-formula approach fundamentally differs from the existing approaches in that the interval partition is the specific means by which reasonable assumptions regard censoring can be formed, and the censoring/death risks are time-updating. While this feature underscores the novelty of our implementation of the g-formula in a broader sense, the nested g-formula cannot be applied to cost data in the absence of repeated outcomes. To our knowledge, the g-computation procedure has not been previously used to estimate parameters involving aggregates of repeated continuous outcomes.

The choice to target either an ITT or a joint causal effect should be scientifically motivated. The joint treatment effect can be realized as a collection of individual treatment effects, rather than the net effect of an intention to treat. We argue that the ITT effect is useful for contextualizing real-world contrasts on a subject-specific level, irrespective of events that occur after treatment commencement. When seeking to make population-level policy recommendations on the sole basis of differences in average cost at the most fundamental level, an approach to estimate the joint causal effect would be more appropriate. Such cost comparisons would be appropriate when presented with treatment options that were comparably safe and effective (or at least, non-inferior in the sense of each not being unacceptably worse than the other when given additional information such as cost and side-effects).

We have empirically demonstrated that the nested g-formula has favorable finite-sample properties including low bias, valid bootstrap standard errors, and proper coverage. Nonrandom subject dropout can be accommodated by the nested g-formula, at least to the extent to which the dropout is explainable by observable covariates. This gives the nested g-formula approach an advantage over alternative approaches that do not account for this commonly encountered limitation. Additionally, our approach showed relatively low sensitivity to departures from distributional assumptions on the outcome in many settings.

One limitation of this approach is that parametric models must be specified to estimate models of interest. This is quite important for cost outcomes which may, in practice, be right skewed. This is of particular relevance when cost distributions differ between groups but are presumed to be the same. However, the parametric assumptions made by the GLMs used in the nested g-formula approach can be tested and separate models can be used for the treatment groups. Moreover, we found that the Normal model was quite flexible even in the presence of heavy right-skewness, suggesting a level of robustness with this simpler approach. A more flexible modeling procedure to account for departures from distributional assumptions in the nested g-formula could be of interest.

The nonparametric bootstrap performed well in representing the true repeat-sample variability. When hypothesis testing is the primary goal, Wald-type tests based on the nonparametric bootstrap standard errors have been suggested to have somewhat lower power than alternatives (\citealp{Davison97}). The parametric bootstrap test has been proposed, whereby a null model is fit and data are repeatedly simulated from the null parameters. The results from those simulated data sets are compared to the unconstrained estimator in question. While this approach could potentially be applicable, it would require specification of the treatment and censoring models, each of which is bypassed in the nested g-formula. Such an approach could also prove computationally taxing. In addition to finding ways to relax distributional and parametric assumptions with more flexible modeling, it would be of interest to compare or develop more optimal hypothesis testing procedures.

\clearpage

\section*{\normalsize Appendix A: Derivation of the parametric nested g-formula.}
We can write the mean potential cost by sequentially invoking the tower property of expectation. Iterating out observations in interval $j - 1$ yields the following expression:
\begin{eqnarray*}
\E[Y_j^{\bar{a}}] & = & \E_{\bar{Y}_{j - 1}, \bar{D}_{j - 1}, \bar{C}_j = 0, \bar{\bL}_j, \bar{A}_j = \bar{a}_j}\left[\E_0[Y_j]\right]\\
~ & = & \E_{\bar{\bL}_{j -1}, \bar{A}_{j - 1} = \bar{a}_{j - 1}, \bar{Y}_{j - 1}, \bar{D}_{j - 1}, \bar{C}_{j - 1} = 0}\left[\E_{1}\left[\E_0[Y_j]\right]\right]\\
~ & = & \E_{\bar{D}_{j - 2}, \bar{C}_{j -1} = 0, \bar{\bL}_{j -1}, \bar{A}_{j - 1} = \bar{a}_{j - 1}, \bar{Y}_{j - 1}}\left[\E_{2}\left[\E_{1}\left[\E_0[Y_j]\right]\right]\right]\\
~ & = & \E_{\bar{Y}_{j - 2}, \bar{D}_{j - 2}, \bar{C}_{j - 1} = 0, \bar{\bL}_{j -1}, \bar{A}_{j - 2} = \bar{a}_{j - 2}}\left[\E_{3}\left[\E_{2}\left[\E_{1}\left[\E_0[Y_j]\right]\right]\right]\right],
\end{eqnarray*}
where we define $\E_{0}[\cdot] \equiv \E_{Y_j}[\cdot|\bar{Y}_{j - 1}, \bar{D}_{j - 1}, \bar{C}_j = 0, \bar{\bL}_j, \bar{A}_j = \bar{a}_j]$, $\E_{1}[\cdot] \equiv \E_{\bL_j}[\cdot|\bar{\bL}_{j - 1}, \bar{A}_{j - 1} = \bar{a}_{j - 1}, \bar{Y}_{j - 1}, \bar{D}_{j - 1}, \bar{C}_{j - 1} = 0]$, $\E_{2}[\cdot] \equiv \E_{D_{j - 1}}[\cdot|\bar{D}_{j - 2}, \bar{C}_{j - 1} = 0, \bar{\bL}_{j - 1}, \bar{A}_{j - 1} = \bar{a}_{j - 1}, \bar{Y}_{j - 1}]$, and $\E_3[\cdot] \equiv \E_{Y_{j - 1}}[\cdot|\bar{Y}_{j - 2}, \bar{D}_{j - 2}, \bar{C}_{j - 1} = 0, \bar{\bL}_{j - 1}, \bar{A}_{j - 2} = \bar{a}_{j - 2}]$. Iteration can be continued in this way sequentially and reduced to the following expression:
\begin{eqnarray*}
\E[Y_j^{\bar{a}}] &=& \iiint \E[Y_j|\bar{\bL}_j = \bar{\bell}_j, \bar{A}_j = \bar{a}_j, \bar{Y}_{j - 1} = \bar{y}_j, \bar{D}_j = \bar{d}_j, \bar{C}_j = 0] \\
~ & ~ & \hspace{0.4in} \times  \hspace{0.03in} \prod_{k = 1}^{j - 1} f_{Y_k|\bar{Y}_{k - 1}, \bar{C}_k, \bar{\bL}_k, \bar{A}_k, \bar{D}_{k - 1}} (y_k, \bar{y}_{k - 1}, 0, \bar{\bell}_k, \bar{a}_k, \bar{d}_{k - 1}) \\
~ & ~ & \hspace{0.4in} \times \hspace{0.03in} \prod_{k = 1}^{j} f_{D_k|\bar{D}_{k - 1}, \bar{Y}_{k}, \bar{C}_{k}, \bar{\bL}_{k}, \bar{A}_{k}} (d_k, \bar{d}_{k - 1}, \bar{y}_{k}, 0, \bar{\bell}_{k}, \bar{a}_{k}) \\
~ & ~ & \hspace{0.4in} \times \hspace{0.03in} \prod_{k = 1}^{j} f_{L_k|\bar{\bL}_{k - 1}, \bar{A}_{k - 1}, \bar{D}_{k - 1}, \bar{Y}_{k - 1}, \bar{C}_{k}} (\bell_k, \bar{\bell}_{k - 1}, \bar{a}_{k - 1}, \bar{d}_{k - 1}, \bar{y}_{k - 1}, 0) d\bar{d}_{j} d\bar{y}_{j} d\bar{\bell}_j.
\end{eqnarray*}
Noting that the conditional expectation of $Y_j$ can be parametrically expressed, we have:
\begin{eqnarray*}
\E[Y_j^{\bar{a}}] &=& \iiint  y_j  \prod_{k = 1}^{j} f_{Y_k|\bar{Y}_{k - 1}, \bar{C}_k, \bar{\bL}_k, \bar{A}_k, \bar{D}_{k - 1}} (y_k, \bar{y}_{k - 1}, 0, \bar{\bell}_k, \bar{a}_k, \bar{d}_{k - 1}) \\
~ & ~ & \hspace{0.4in} \times \hspace{0.03in} \prod_{k = 1}^{j} f_{D_k|\bar{D}_{k - 1}, \bar{Y}_{k}, \bar{C}_{k}, \bar{\bL}_{k}, \bar{A}_{k}} (d_k, \bar{d}_{k - 1}, \bar{y}_{k}, 0, \bar{\bell}_{k}, \bar{a}_{k}) \\
~ & ~ & \hspace{0.4in} \times \hspace{0.01in} \prod_{k = 1}^{j} f_{L_k|\bar{\bL}_{k - 1}, \bar{A}_{k - 1}, \bar{D}_{k - 1}, \bar{Y}_{k - 1}, \bar{C}_{k}} (\bell_k, \bar{\bell}_{k - 1}, \bar{a}_{k - 1}, \bar{d}_{k - 1}, \bar{y}_{k - 1}, 0) d\bar{d}_{j} d\bar{y}_{j} d\bar{\bell}_j.
\end{eqnarray*}
Writing $\E[\mY^{\bar{a}}] = \sum_{j = 1}^J \E[Y_j^{\bar{a}}]$, we have the desired result.

\clearpage

\section*{\normalsize Appendix B: Summary of fixed simulation parameters.}

\begin{table}[h!]
\centering
\begin{tabular}{rlc}
\hline
Scenario & Model Description & Parameter Value \\ \hline
$({\alpha_{0, 1}}, \sigma_{L_1}^2)$ & Baseline confounder & $(0, 1)$  \\
$(\eta_{0, 1}, \eta_{1, 1})$ & Baseline treatment probability & $(0, 1)$ \\
$(\beta_{0, 1}, \beta_{1, 1}, \beta_{2, 1})$ & Baseline mean cost & $(20, 1, 2)$ \\ 
$(\gamma_{0, 1}, \gamma_{1, 1}, \gamma_{2, 1})$ & Death probability after first interval & $(-5, 0.3, 0.05)$\\ 
$(\alpha_0, \dots, \alpha_3, \sigma_{L}^2)$ & Follow-up confounder & $(-1.09, 0.5, 0.5, 0.04, 4)$\\
$(\eta_{0}, \dots, \eta_4)$ & Follow-up treatment probability & $(-1.34, 0.4, 0.6, 1, 0.04)$ \\
$(\beta_{0}, \dots, \beta_{5})$ & Follow-up mean cost & $(10.65, 0.2, 0.4, 0.2, 0.4, 0.05)$ \\ 
$(\gamma_0, \dots, \gamma_3)$ & Follow-up probability of death & $(-3, 0.1, 0.2, 0.03)$ \\ 
\hline
\end{tabular}
\end{table}
%
%
%

\clearpage

\section*{\normalsize Appendix C: Review of Inverse-Weighting Approaches for ITT Effects}

We review the weighting approach of \citet{Lin00} and the IPTW-based extension discussed by \citet{Li16} to address censoring in the analysis of cost data.

\subsection*{\normalsize C.1: Notation}

Let $i = 1, \dots, N$ index study subjects, and let $0 \equiv \tau_0 < \tau_1 < \cdots < \tau_{J - 1} < \tau_J \equiv \tau$ denote a discrete partition of the time interval of interest. Again let $Y_{ij}$ denote the cost in interval $[\tau_{j - 1}, \tau_j)$, and $\mY_i = \sum_{j = 1}^{J} Y_{ij}$ the cumulative cost over $[0, \tau]$. Define $A_{i}$ to be baseline treatment, and $\bL_{i}$ the set of baseline confounders for subject $i$. Let $T_i^C$ and $T_i^D$ denote the censoring and survival times for subject $i$, respectively, and $T_i = \min\lbrace T_i^D, \tau \rbrace$. Further let $\delta_i = 1(T_i^C \geq T_i)$ denote the indicator of observing complete cost data for subject $i$. Finally, let $X_i = \min \lbrace T_i^D, T_i^C \rbrace$, and $\tilde{\delta}_i = 1(T_i^C \geq T_i^D)$ the indicator of observing death for subject $i$.

\subsection*{\normalsize C.2: The partitioned IPCW estimator}

\citet{Lin00} proposed inverse-weighted equation to estimate the intent-to-treat effect $\Delta_{\text{ITT}} \equiv \E[\mY_i|A_i = 1, \bL_i] - \E[\mY_i|A_i = 0, \bL_i]$. Let $\widehat{K}(t)$ denote the Kaplan-Meier estimator of $K(t) = P(T^C > t)$ based on $\lbrace (X_i, 1 - \tilde{\delta}_i) : i = 1, \dots, N \rbrace$. A $Z$-estimator $\widehat{\Delta}$ can be obtained through the following estimating equations:
\begin{eqnarray}
\sum_{i|\delta_i = 1} \frac{\bW_i(\mY_i - \bW_i^T\zeta)}{\widehat{K}(T_i)} = 0,
\end{eqnarray}
where $\bW_i = (1, A_i, \bL_i)^T$, and $\zeta = (\mu_0, \Delta_{\text{ITT}}, \zeta_1)^T$; here, $\zeta_1$ corresponds to confounders. Only complete cases contribute to the estimating equations (all subjects contribute to the weights).

Recognizing the problem of efficiency loss associated with heavy censoring, \citet{Lin00} then proposed the use of interval cost information to allow censored subjects to contribute to the estimation of $\Delta_{\text{ITT}}$ for as many intervals as they are not censored. In particular, let $T_{ij} = \min(T_i^D, \tau_j)$ and $\delta_{ij} = 1(T_i^C \geq T_{ij})$. Then $\Delta_j$, the difference in mean cost between the two treatment groups within interval $j$, can be estimated from the following estimating equations, for $j = 1, \dots, J$:
\begin{eqnarray}
\sum_{i|\delta_{ij} = 1} \frac{\bW_i(Y_{ij} - \bW_i^T\zeta_j)}{\widehat{K}(T_{ij})} = 0,
\end{eqnarray}
where $\zeta_j = (\mu_{0j}, \Delta_{\text{ITT}}^{(j)}, \zeta_{1j})^T$. Using the resulting estimators from these $J$ estimating equations, we can estimate the difference in mean total cost by $\widehat{\Delta}_{\text{ITT}} = \sum_{j = 1}^J \widehat{\Delta}_{\text{ITT}}^{(j)}$.

These estimating equations result in closed-form expressions for $\widehat{\Delta}$ and enjoy both consistency and asymptotic normality under suitable regularity conditions. Sandwich-based variance estimators suitable for weighted $Z$-estimators can be used to construct confidence intervals and conduct inference. Of note, the Cox Proportional Hazards model can be used to estimate $\widehat{K}(t)$ if one wishes to condition on baseline observable covariates.

\subsection*{\normalsize C.3: Propensity score adjustment}

\citet{Li16} proposed (among other forms of propensity score adjustment) an extension to include IPTW-based weights. Confounding can therefore be addressed without altering the target of inference (i.e., without needing a conditional model for mean cost). Let $\pi_i = \pi_i(\bL_{i}) = P(A_i = 1|\bL_{i})$; it is standard to estimate propensity scores $\pi_i$ by obtaining predictions from a logistic regression  model. First, an estimate of $(\psi_0, \boldsymbol{\psi})$ is obtained from the model $\logit(\pi_i) = \psi_0 + \bL_{i}^T\boldsymbol{\psi}$ using, for example, maximum likelihood. In turn, predicted probabilities can be obtained: $\hat{\pi}_i = \exp(\hat{\psi}_0 + \bL_{i}^T\hat{\boldsymbol{\psi}})/[1 + \exp(\hat{\psi}_0 + \bL_{i}^T\hat{\boldsymbol{\psi}})]$.

Weights can be incorporated into the equations of the IPCW estimator to estimate $\Delta_{\text{ITT}}$. For example, $\widehat{\Delta}_{\text{ITT}}^{(j)}$ from the partition approach could be obtained by solving the estimating equations
\begin{eqnarray}
\sum_{i|\delta_{ij} = 1} \frac{\bW_i(Y_{ij} - \bW_i^T\zeta_j)}{\widehat{K}(T_{ij})} \cdot \left[\frac{A_i}{\hat{\pi_i}} + \frac{1 - A_i}{1 - \hat{\pi_i}}\right] = 0.
\end{eqnarray}
The estimator $\widehat{\Delta}_{\text{ITT}}$ has a closed-form expression and, too, is asymptotically normal and consistent under weight positivity and correct propensity score model specification.

\clearpage

\bibliography{Spieker_ArXiv}


\end{document}